\documentstyle[preprint,aps,epsf]{revtex}                                       
\begin{document}

\tighten
\draft
\preprint{
\vbox{
\hbox{August 1996}
\hbox{U.MD. PP\# 97-16}
\hbox{DOE/ER/40762-095}
}}

\title{Strangeness in the Nucleon on the Light-Cone}
\author{W. Melnitchouk and
	M. Malheiro\thanks{
Permanent address: 
Instituto de F\'\i sica, 
Universidade Federal Fluminense, 
24210-340, Niter\'oi, R. J., Brazil. 
Partially supported by CAPES of Brazil, no. BEX1278/95-2}}
\address{Department of Physics, 
	University of Maryland, 
	College Park, Maryland 20742-4111}

\maketitle

\begin{abstract}
Strange matrix elements of the nucleon are calculated within 
the light-cone formulation of the meson cloud model.
The $Q^2$ dependence of the strange vector and axial vector 
form factors is computed, and the strangeness radius and
magnetic moment extracted, both of which are found to be 
very small and slightly negative.
Within the same framework one finds a small but non-zero excess
of the antistrange distribution over the strange at large $x$.
Kaon loops are unlikely, however, to be the source of a large 
polarized strange quark distribution.
\end{abstract}
\pacs{PACS numbers: 13.60.Hb, 13.87.Fh, 13.88.+e
\vspace*{4cm}	\\
To be published in Phys. Rev. C (Jan 1997)}

\section{Introduction}

There has been considerable discussion recently about strange matrix 
elements of the nucleon \cite{REP}.
Much impetus for this was generated by the deep-inelastic scattering 
experiments with polarized targets at CERN and more recently at SLAC 
\cite{SPIN}, which seemed to imply a large polarized strange quark 
distribution in the proton.
At about the same time a measurement of the elastic neutrino--proton 
scattering cross section \cite{ELNU} at lower values of $Q^2$ had also 
suggested a non-zero value for the strange axial vector form factor 
of the proton \cite{GARV}.

These observations spurred many investigations of other processes 
in which traces of strangeness in the nucleon could be detected.
It was argued \cite{KM} for example that semi-leptonic neutral current
scattering experiments could be used to extract the strange vector
as well as axial vector form factors. 
Parity-violating electron scattering experiments were proposed
as a means of probing neutral current form factors \cite{PARITY}. 
Approved parity-violating experiments at MIT-Bates \cite{BATES} 
and Jefferson Lab \cite{JEFF} will provide information on the 
strangeness form factors at low and intermediate $Q^2$ values, and
more precisely determine the strange radius and magnetic moment of 
the nucleon.
At higher $Q^2$, the CCFR Collaboration \cite{CCFR} has recently 
investigated a possible asymmetry between the strange and antistrange
quark distributions in neutrino deep-inelastic scattering.
Perturbative QCD alone would be expected to produce identical $s$ and 
$\overline s$ distributions, while any asymmetry would imply the presence 
of non-perturbative effects in the nucleon at deep-inelastic scales.

Of course the strangeness content of the nucleon is not a scale-invariant
concept --- strange matrix elements are not renormalization group invariant.
The evolution with $Q^2$ of the moments of the polarized deep-inelastic
strange quark distribution can be studied using the Altarelli-Parisi 
equations, so that a zero value for $\Delta s$ at a low scale is not 
necessarily incompatible with a non-zero value at a higher scale 
\cite{EVOL}.
Naturally, extending such analyses to very low values of $Q^2$ is
problematic --- at some stage non-perturbative techniques or models
of QCD need to be invoked when describing data.
Recently Tang and Ji \cite{JI} attempted to link the strangeness radius 
of the nucleon with the densities of strange and antistrange deep-inelastic 
quark distributions, arguing on the basis of quark ``locality'' that the 
distributions in coordinate space and momentum space should be correlated.
In order to investigate this proposal more quantitatively, we will explore
this relationship within a simple model, namely the meson cloud model.

In the meson cloud model, the strangeness of the nucleon is assumed to 
be carried by the kaon--hyperon components of the physical nucleon.
This model has in the past been utilized to calculate corrections to low 
energy nucleon properties \cite{HOL,KHP,DM,ZOL,MB,MD,ITO}, and has 
also been invoked to describe flavor symmetry breaking in nucleon sea 
quark distributions \cite{AWT,ST,MTV,JUL}.
However, direct comparison of results for the strange matrix elements 
in meson cloud models has thus far not been possible due to the fact 
that form factors and structure functions are usually calculated within 
different theoretical frameworks.
While the natural framework for analyzing strange quark distributions 
is the light-cone \cite{MTV,JUL}, strange form factor calculations have 
usually been performed within covariant perturbation theory in instant-form 
quantization \cite{HOL,KHP,MB,MD,ITO}.
(A light-cone constituent quark model was used in Ref.\cite{ITO} to 
obtain the coupling of the photon to the kaon--hyperon cloud of the 
nucleon, however the kaon dressing itself was not described in terms 
of light-cone dynamics.)
Meaningful comparison of meson cloud corrections to light-cone quark 
distributions and low-$Q^2$ form factors obviously requires a single 
framework to be used for both.
In the present work we shall present for the first time a light-cone 
analysis of the strange nucleon form factors, using the same framework 
in which structure functions are calculated \cite{MTV,JUL}.
We believe this is the only systematic way to test the hypothesis of 
Ref.\cite{JI} that the properties of the nucleon sea at low and high
$Q^2$ might somehow be correlated.

In Sec.II we briefly outline the pertinent features of the meson 
cloud model on the light-cone.
More detailed accounts can be found in Ref.\cite{MTV}.
We explain that the advantage of this framework lies in that it allows 
for a more consistent treatment of the hadronic meson--baryon vertex 
functions. 
Unlike previous calculations within covariant perturbation theory, 
which have usually had to make {\em ad hoc} assumptions about the 
dependence of the vertex functions on various loop momenta, one can 
formally avoid this on the light-cone, while simultaneously satisfying 
charge and momentum conservation.
The kaon cloud model is applied in Sec.III to the calculation of the 
$Q^2$ dependence of the strange vector and axial vector form factors 
of the nucleon, from which the strangeness radius, strange magnetic 
moment and strange axial charge are extracted.
Sec.IV discusses the contribution from the kaon cloud to the deep-inelastic 
strange quark distribution, and to a possible strange--antistrange asymmetry
in the nucleon.
We find that the asymmetry is consistent with the latest neutrino 
deep-inelastic scattering data, within experimental errors.
We also estimate the contribution to the polarized strange quark
distribution from kaon loops.
Finally our findings are summarized in Sec.V.

\section{Light-Cone Meson Cloud Model of the Nucleon} 

The basic hypothesis of the meson cloud model of the nucleon is that
the nucleon on the light-cone has internal meson and baryon degrees 
of freedom.
The physical nucleon state (momentum $P$) can then be expanded 
(in the one-meson approximation) as a series involving bare nucleon 
and two-particle meson--baryon states: 
\begin{eqnarray}
\label{Fock}
\left| N(P) \rangle_{\rm phys} \right.
&=& \sqrt{Z}\
\left\{ \left| N(P) \rangle_{\rm bare} \right. \right.       \nonumber\\
& & +\  \sum_{B,M} \int dy\ d^2{\bf k}_\perp\
     g_{MNB}\ \phi_{BM}(y,{\bf k}_\perp)\
     \left| B (y,{\bf k}_\perp); M (1-y,-{\bf k}_\perp) \rangle \right.
\left.
\right\}.
\end{eqnarray}
The function $\phi_{BM}(y,{\bf k}_\perp)$ is the probability amplitude for 
the physical nucleon $N$ to be in a state consisting of a baryon $B$ and 
meson $M$, having transverse momenta ${\bf k}_\perp$ and $-{\bf k}_\perp$,
and carrying longitudinal momentum fractions $y = k_+/P_+$ and 
1--$y = (P_+ - k_+)/P_+$, respectively. 
The bare nucleon probability is denoted by $Z$, and $g_{MNB}$ is the $MNB$ 
coupling constant.
The one-meson approximation in Eq.(\ref{Fock}) is valid as long as the 
meson cloud is relatively soft ($Z \alt 1$).
It will progressively break down for harder $MNB$ vertex functions, 
at which point one will need to include two-meson and higher order 
Fock state components in Eq.(\ref{Fock}).

The strangeness of the nucleon in this model is carried by the
$\left| \Lambda; K \rangle \right.$, $\left| \Sigma; K \rangle \right.$, 
etc. Fock state components, and we shall consider only these henceforth.
Therefore the properties of the bare nucleon state (such as its
intrinsic form factor, or its structure function) will not be
relevant for considerations of the nucleon strangeness content.
For the sake of presentation we will consider the $\Lambda$ hyperon
as representative in our discussions, although contributions from the 
$\Sigma$ will be included in the numerical results in Secs.III \& IV. 
All expressions for the $K\Sigma$ distributions can be obtained from 
the $K\Lambda$ by appropriate replacement of masses, coupling constants
and isospin factors.
Contributions from heavier mesons and hyperons, such as $K^*$, $Y^*$,
etc. can easily be added, although their role is insignificant as long
as the $MNB$ vertex functions are not very hard \cite{MTV}.

According to Eq.(\ref{Fock}), the probability to find a $\Lambda$ inside 
a nucleon with light-cone momentum fraction $y$ is given by the 
$N \rightarrow K\Lambda$ splitting function (to leading order in
the coupling constant):
\begin{eqnarray}
f_{\Lambda K}(y) 
&=& g_{KN\Lambda}^2 \int d^2{\bf k}_\perp\
	\left| \phi_{\Lambda K}(y,{\bf k}_\perp) \right|^2.
\end{eqnarray}
Charge and momentum conservation require that this must also be the
probability to find a kaon inside a nucleon with light-cone momentum 
fraction $1-y$.
The kaon distribution function, $f_{K \Lambda}(y)$, should therefore
be related to the baryon distribution function by
\begin{eqnarray}
\label{sym} 
f_{K\Lambda}(y) &=& f_{\Lambda K}(1-y),
\end{eqnarray}
for all $y$, if the above interpretation is valid.
This guarantees equal numbers of mesons emitted by the nucleon, 
$\langle n \rangle_{K\Lambda} = \int_0^1 dy\ f_{K\Lambda}(y)$,
and virtual baryons accompanying them,
$\langle n \rangle_{\Lambda K} = \int_0^1 dy\ f_{\Lambda K}(y)$:
\begin{mathletters}
\label{cons}
\begin{eqnarray}
\label{Ncons}
\langle n \rangle_{K\Lambda} &=& \langle n \rangle_{\Lambda K}.
\end{eqnarray}
This is just the statement that the nucleon has zero net strangeness.
Momentum conservation imposes the further requirement that
\begin{eqnarray}
\label{Mcons} 
\langle y \rangle_{K\Lambda}\ +\ \langle y \rangle_{\Lambda K}
&=& \langle n \rangle_{K\Lambda}, 
\end{eqnarray}
\end{mathletters}%
where
$\langle y \rangle_{K\Lambda} = \int_0^1 dy\ y\ f_{K\Lambda}(y)$ and
$\langle y \rangle_{\Lambda K} = \int_0^1 dy\ y\ f_{\Lambda K}(y)$
are the average momentum fractions carried by kaon and $\Lambda$, 
respectively.
Equations (\ref{Ncons}) and (\ref{Mcons}), and in fact similar relations
for all higher moments of $f(y)$, follow automatically from Eq.(\ref{sym}).
Any consistent treatment of the meson cloud model must therefore reproduce
this symmetry relation.

The advantage of a light-cone formulation \cite{DM,ZOL,MTV,JUL,BROD} 
is that one can construct a covariant framework in which Eq.(\ref{sym}) 
is manifestly preserved, regardless of the dynamics of the meson--baryon
system.
If the dynamics are parametrized in terms of an effective hadronic 
meson--baryon vertex function, the symmetry relations 
(\ref{sym})--(\ref{cons}) can be satisfied if the vertex functions
are functions of the invariant mass squared, ${\cal M}^2$, of the 
$K\Lambda$ system:
\begin{eqnarray}
\label{M2}
{\cal M}^2\ =\ (p_K + p_\Lambda)^2
	   &=& {k_\perp^2 + M_{\Lambda}^2 \over y}
	    +  {k_\perp^2 + m_K^2 \over 1-y},
\end{eqnarray}
where $M_{\Lambda}$ and $m_K$ are the masses of $\Lambda$ and kaon,
respectively, and $p_K$ and $p_\Lambda$ their four-momenta.
Note that the variable ${\cal M}^2$ is related to the virtualities of 
the kaon ($t = (p_\Lambda-p_N)^2$) and $\Lambda$ ($u = (p_K-p_N)^2$) 
by:
\begin{eqnarray}
\label{stu}
{\cal M}^2 + t + u &=& M^2 + m_K^2 + M_{\Lambda}^2,
\end{eqnarray}
where $t = -( k_\perp^2 + (1-y) (M_{\Lambda}^2 - y M^2) ) / y$ 
and $u = -( k_\perp^2 + y (m_K^2 - (1-y) M^2) ) / (1-y)$.
A hadronic vertex function which is a function of ${\cal M}^2$ will 
automatically have the correct crossing symmetry structure in the 
$t$ and $u$ channels \cite{ZOLL,STU} (see Eqs.(\ref{FF}), (\ref{FFt})
below).

In instant-form calculations, on the other hand, the functions $f(y)$ 
are usually evaluated in terms of $KN\Lambda$ vertex functions taken 
from fits to $N\Lambda \rightarrow N\Lambda$ scattering data \cite{HOLZ},
where the vertex functions are assumed to depend only on the virtuality
$t$ of the off-shell kaon.
In this case the $t \leftrightarrow u$ symmetry is lost, and the relation 
(\ref{sym}) cannot be maintained, making verification of the charge and 
momentum conservation relations, Eqs.(\ref{cons}), very difficult 
\cite{MTV,MST}.
Actually, the non-preservation of the relation (\ref{sym}) is not surprising
since the splitting functions $f(y)$ refer to probability distributions of 
specific {\em particles}, and a probabilistic interpretation does not hold
in all reference frames (since particle number is not preserved by Lorentz
boosts).
Indeed, for structured particles the factorization of the $\gamma^* N$ 
cross section itself into $\gamma^* K$ and $K N$ (or $\gamma^* \Lambda$ 
and $\Lambda N$) cross sections is not valid in all frames of reference 
\cite{MST}.
Such factorization, or convolution, can in fact only be achieved by
eliminating antiparticle degrees of freedom, which can formally be done 
only in the infinite momentum frame \cite{IMF}, or on the light-cone.

\section{Strange Form Factors}

We will be interested in the strange matrix elements 
$\langle N | \overline s \Gamma s | N \rangle$ of operators $\Gamma$,
where $\Gamma = \gamma_{\mu}$ or $\gamma_5 \gamma_{\mu}$.
While these have been investigated in earlier studies within the meson 
cloud model \cite{HOL,KHP,MB,MD,ITO}, we will present here the first 
analysis of the strange form factors on the light-cone, 
in which the distribution functions of kaons and hyperons in the nucleon 
have the correct symmetry properties as specified in Eq.(\ref{sym}).
This will be our main contribution to the form factor discussion.
Other aspects of the problem, such as the modeling of the intrinsic 
$K$ and $\Lambda$ form factors, or the truncation of the Fock state
expansion in Eq.(\ref{Fock}), are not developed here beyond what
exists already in the literature.

The momentum dependence of the matrix elements of the strange vector 
current $J_{\mu}^s = \overline s \gamma_{\mu} s$ is conventionally 
written:
\begin{eqnarray}
\langle N(P') | J_{\mu}^s(0) | N(P) \rangle
&=& \overline u(P')
\left( \gamma_{\mu} F_1^s(Q^2)
     + {i\sigma_{\mu\nu} q^{\nu} \over 2 M} F_2^s(Q^2)
\right) u(P), 
\end{eqnarray}
where $P$ and $P'$ are the initial and final nucleon momenta, 
and $q = P - P'$ is the momentum transferred to the nucleon.
The form factors $F_1^s$ and $F_2^s$ are the Dirac and Pauli 
strange form factors, and $u(P)$ is the free Dirac spinor of 
the nucleon.
For convenience we work in a Lorentz frame (viz. the Breit frame)
where the photon momentum $q$ is purely transverse, so that 
$q^2 = -q_\perp^2 \equiv -Q^2$.
The various particle momenta are parameterized as in Ref.\cite{BGP}.
Zero net strangeness in the nucleon requires the vanishing of the 
Dirac form factor at the on-shell photon point, $F_1^s(0) = 0$. 
On the light-cone the form factors can be evaluated by choosing 
the $\mu=+$ (or ``good'') component of the current $J_{\mu}^s$ 
\cite{BROD}.

Usually one introduces more convenient combinations of $F_{1,2}^s$,
in the form of the Sachs electric ($G_E^s$) and magnetic ($G_M^s$) 
form factors: 
\begin{mathletters}
\begin{eqnarray}
G_E^s(Q^2)
&=& F_1^s(Q^2) - {Q^2 \over 4 M^2} F_2^s(Q^2),	\\
G_M^s(Q^2)
&=& F_1^s(Q^2) + F_2^s(Q^2). 
\end{eqnarray}
\end{mathletters}%
The conventional definition of the strangeness radius of the nucleon is 
in terms of $G_E^s$:
\begin{mathletters}
\begin{eqnarray}
r_s^2
&\equiv& \int d^3r 
\langle N | \overline s\ {\bf r}^2 \gamma_0\ s | N \rangle      \\
&=& - 6 \left.{d G_E^s(Q^2) \over dQ^2}\right|_{Q^2=0}.
\end{eqnarray}
\end{mathletters}%
An alternative definition of the strangeness radius is in terms of the 
Dirac form factor:
\begin{eqnarray}
r_{s, Dirac}^2
&=& - 6 \left.{d F_1^s(Q^2) \over dQ^2}\right|_{Q^2=0}.
\end{eqnarray}
The strange magnetic moment of the nucleon is defined by:
\begin{mathletters}
\begin{eqnarray}
\mu_s 
&\equiv& {1 \over 2} \int d^3r
\langle N | ({\bf r} \times \overline s \vec\gamma s)_z | N \rangle\\ 
&=& F_2^s(0)\ =\ G_M^s(0).
\end{eqnarray}
\end{mathletters}%

The contributions of the kaon cloud to the strange form factors are 
represented in Fig.1.
To maintain the analogy with the structure function calculations in 
Sec.IV, we will present the results for $F_{1,2}^s$, and evaluate 
$G_{E,M}^s$ from these.
Therefore we define:
\begin{eqnarray}
F_i^s(Q^2)
&=& F_i^{(\Lambda)}(Q^2)\ +\ F_i^{(K)}(Q^2),
\ \ \ \ \ \ \ \ i=1,2, 
\end{eqnarray}
where $F_i^{(\Lambda)}$ and $F_i^{(K)}(Q^2)$ refer to the $\Lambda$ 
and $K$ interaction diagrams in Figs.1(a) and (b), respectively.
The $\Lambda$ contribution to $F_i^s$ can be written:
\begin{eqnarray}
\label{FiLam}
F_i^{(\Lambda)}(Q^2)
&=& Q_{\Lambda} \int_0^1 dy\ f_i^{(\Lambda)}(y,Q^2)\ H_{\Lambda}(Q^2), 
\ \ \ \ \ \ \ \ i=1,2, 
\end{eqnarray}
where the $\Lambda$ light-cone distribution functions $f_i^{(\Lambda)}$ 
are:
\begin{mathletters}
\label{f12L}
\begin{eqnarray}
\label{f1L}
f_1^{(\Lambda)}(y,Q^2)
&=&
{ g_{KN\Lambda}^2 \over 16 \pi^3 }
\int { d^2{\bf k}_\perp \over y^2 (1-y) }
{ {\cal F}({\cal M}^2_{\Lambda K,i})\
  {\cal F}({\cal M}^2_{\Lambda K,f})
  \over ({\cal M}^2_{\Lambda K,i} - M^2)
	({\cal M}^2_{\Lambda K,f} - M^2) }
\nonumber\\
& & \times
\left(
  k_\perp^2 + (M_{\Lambda} - y M)^2
- (1-y)^2 { q_\perp^2 \over 4 }
\right),                                        \\
\label{f2L}
f_2^{(\Lambda)}(y,Q^2)
&=&
{ g_{KN\Lambda}^2 \over 16 \pi^3 }
\int { d^2{\bf k}_\perp \over y^2 (1-y) }
{ {\cal F}({\cal M}^2_{\Lambda K,i})\
  {\cal F}({\cal M}^2_{\Lambda K,f})
  \over ({\cal M}^2_{\Lambda K,i} - M^2)
	({\cal M}^2_{\Lambda K,f} - M^2) }
\nonumber\\
& & \times
(-2 M) (1-y) (M_{\Lambda} - y M).
\end{eqnarray}
\end{mathletters}%
For the $KN\Lambda$ vertex we assume a pseudoscalar $i\gamma_5$ interaction 
(the same results are obtained with a pseudovector coupling), with 
$g_{KN\Lambda}$ the coupling constant and ${\cal F}({\cal M}^2_{\Lambda K})$ 
the hadronic vertex function.
Note that identical expressions to these can be obtained using 
time-ordered perturbation theory in the infinite momentum frame
\cite{MTV,IMF}.
For the sign of the strangeness we adopt the convention of Jaffe \cite{JAF},
so that $Q_{\Lambda} = +1$ is the strangeness charge of the $\Lambda$.
The intrinsic $\Lambda$ form factor $H_{\Lambda}(Q^2)$ contains possible
$Q^2$ dependence in the $\gamma^* \Lambda$ interaction vertex.
The squared center of mass energies in Eqs.(\ref{f12L}) are:
\begin{mathletters}
\begin{eqnarray}
{\cal M}^2_{\Lambda K,i}
&=& {\cal M}^2\
 +\ {{\bf q}_\perp \over y} \cdot
    \left( (1-y) {{\bf q}_\perp \over 4} + {\bf k}_\perp \right),\\
{\cal M}^2_{\Lambda K,f}
&=& {\cal M}^2\
 +\ {{\bf q}_\perp \over y} \cdot
    \left( (1-y) {{\bf q}_\perp \over 4} - {\bf k}_\perp \right),
\end{eqnarray}
\end{mathletters}%
with ${\cal M}^2$ defined in Eq.(\ref{M2}).

The contributions to $F_{1,2}^s$ from the coupling to the kaon 
in Fig.1(b) are written:
\begin{eqnarray}
\label{FiK}
F_i^{(K)}(Q^2) 
&=& Q_K \int_0^1 dy\ f_i^{(K)}(y,Q^2)\ H_K(Q^2), 
\ \ \ \ \ \ \ \ i=1,2, 
\end{eqnarray}
where $Q_K = -1$ is defined to be the strangeness charge of the kaon, 
and where the kaon light-cone distribution functions $f_{1,2}^{(K)}$ 
are:
\begin{mathletters}
\label{f12K}
\begin{eqnarray}
\label{f1K} 
f_1^{(K)}(y,Q^2) 
&=&
{ g_{KN\Lambda}^2 \over 16 \pi^3 }
\int { d^2{\bf k}_\perp \over y^2 (1-y) }
{ {\cal F}({\cal M}^2_{K\Lambda,i})\ 
  {\cal F}({\cal M}^2_{K\Lambda,f})
  \over ({\cal M}^2_{K\Lambda,i} - M^2)
	({\cal M}^2_{K\Lambda,f} - M^2) } 
\nonumber \\
& & \times
\left( 
  k_\perp^2 + (M_{\Lambda} - y M)^2 
- y^2 {q_\perp^2 \over 4} 
\right),                         \\
\label{f2K}
f_2^{(K)}(y,Q^2)
&=&
{ g_{KN\Lambda}^2 \over 16 \pi^3 }
\int { d^2{\bf k}_\perp \over y^2 (1-y) }
{ {\cal F}({\cal M}^2_{K\Lambda,i})\
  {\cal F}({\cal M}^2_{K\Lambda,f})
  \over ({\cal M}^2_{K\Lambda,i} - M^2)
	({\cal M}^2_{K\Lambda,f} - M^2) }
\nonumber \\
& & \times
\left(
2 M y (M_{\Lambda} - y M)
\right). 
\end{eqnarray}
\end{mathletters}%
The $K\Lambda$ squared center of mass energies are: 
\begin{mathletters}
\begin{eqnarray}
{\cal M}^2_{K\Lambda,i}
&=& {\cal M}^2\ 
 +\ {{\bf q}_\perp \over 1-y} \cdot 
    \left( y\ {{\bf q}_\perp \over 4} + {\bf k}_\perp \right),	\\ 
{\cal M}^2_{K\Lambda,f}
&=& {\cal M}^2\ 
 +\ {{\bf q}_\perp \over 1-y} \cdot 
    \left( y\ {{\bf q}_\perp \over 4} - {\bf k}_\perp \right). 
\end{eqnarray}
\end{mathletters}%

As discussed in Sec.II, on the light-cone it is natural to take the 
$KN\Lambda$ vertex function to be a function of ${\cal M}_{K\Lambda}^2$, 
which guarantees local gauge invariance \cite{ZOL,ZOLL} as well as 
energy-momentum conservation, as embodied in Eq.(\ref{sym}).
In instant-form approaches gauge invariance has sometimes been enforced
through introduction of seagull diagrams \cite{MB,MD,ITO}, which can be
generated according to various prescriptions \cite{WT} in order to
satisfy the Ward-Takashi identity.
However, as observed by Musolf and Burkardt \cite{MB}, such prescriptions
are not unique, since additional seagull terms which individually satisfy 
the Ward-Takahashi identity are also allowed.
Furthermore, inconsistencies with some of the prescriptions, when applied 
to loop calculations, have also recently been pointed out by Wang and 
Banerjee \cite{WB}.

With ${\cal M}^2$ dependent vertex functions the functions $f_1^{(K)}$ 
and $f_1^{(\Lambda)}$ in Eqs.(\ref{f1K}) and (\ref{f1L}) at $Q^2=0$ 
satisfy the relation (\ref{sym}) explicitly, which they must because of 
strangeness conservation.
The functions $f_2^{(K)}$ and $f_2^{(\Lambda)}$ at $Q^2=0$, on the other
hand, do not need to satisfy such a relation, because they have their
origin in the different spin couplings of the $\gamma^*$ to the spin-0 $K$
and spin-1/2 $\Lambda$.
Indeed, the contribution to the strange magnetic moment of the nucleon
would be zero if $f_2^{(K,\Lambda)}$ exhibited such a symmetry.
{}From Eqs.(\ref{f2L}) and (\ref{f2K}) one can see in fact that $\mu_s$
in the kaon cloud model is always $< 0$.

Although it is clear that in order to preserve the relation (\ref{sym})
the vertex function should be a function of the $K\Lambda$ invariant
mass, its specific functional dependence is not prescribed, and several 
forms have been suggested in the literature \cite{MTV,JUL,BGP}.
In our numerical studies we will use a simple monopole-type function:
\begin{eqnarray}
\label{FF}
{\cal F}({\cal M}_{K\Lambda}^2)
&=&
\left( { \Lambda_{K\Lambda}^2 + M^2 \over
         \Lambda_{K\Lambda}^2 + {\cal M}^2_{K\Lambda} }
\right),
\end{eqnarray}
where $\Lambda_{K\Lambda}$ is the cut-off mass parameter.
In the light-cone formulation the cut-off directly determines the 
probability of finding the physical nucleon in a $K\Lambda$ configuration,
which provides a constraint on the range of $\Lambda_{K\Lambda}$ for which
the average kaon number density can be viewed as reliable. 
(In covariant instant-form calculations the probability for each 
individual Fock state component is frame-dependent, since particle 
number is not invariant under Lorentz boosts.)
For values larger than $\sim 1.5$ GeV the average probability to find 
kaons in the nucleon would be $\langle n \rangle_{K\Lambda} \agt 10\%$, 
so that without including higher Fock state components, the one-meson 
approximation in Eq.(\ref{Fock}) could not be considered trustworthy.
Therefore cut-off masses between $\Lambda_{K\Lambda} = 0.7$ and $1.3$ GeV,
for which $\langle n \rangle_{K\Lambda} \approx$ 3--7\%, 
can be considered as representative of a reasonable range for which 
the model (\ref{Fock}) is a valid approximation (for these values 
the corresponding probability to find pions in the nucleon is 
$\sim$ 30\%--50\%).

On the other hand, one can also try to constrain $\Lambda_{K\Lambda}$ 
phenomenologically by fitting, within the kaon cloud model, the available
strange quark distribution data \cite{ST,MTV,FMS}, as well as the inclusive
$pp \rightarrow \Lambda X$ production data \cite{INCLAM}.
For the monopole parametrization (\ref{FF}) one typically finds values
of $\Lambda_{K\Lambda} \alt 1.3$ GeV, with larger cut-offs difficult to
accommodate \cite{MTV} (see also Sec.IV).
In instant-form calculations \cite{MB,MD,ITO} one has often tried to 
connect the vertex function cut-offs with those obtained from $N\Lambda$
potential model fits \cite{HOLZ}, where the vertex function is assumed 
to depend on the virtuality $t$ of the kaon, 
\begin{eqnarray}
\label{FFt}
{\cal F} &=&
\left( { \widetilde\Lambda_{K\Lambda}^2 - m_K^2 \over
         \widetilde\Lambda_{K\Lambda}^2 - t }
\right).
\end{eqnarray}
Here the vertex function cut-offs are found to be typically 
$\widetilde\Lambda_{K\Lambda} \sim$ 1.2--1.5 GeV.
Note that the form (\ref{FFt}) is an approximation to the 
${\cal M}^2$ dependent light-cone vertex function in Eq.(\ref{FF}),
obtained by taking $u \rightarrow M_{\Lambda}^2$ and 
$\widetilde\Lambda_{K\Lambda}^2 
\rightarrow \Lambda_{K\Lambda}^2 + M^2 + m_K^2$,
see Eq.(\ref{stu}).
A typical value of $\widetilde\Lambda_{K\Lambda} \sim 1.4$~GeV
would therefore correspond to $\Lambda_{K\Lambda} \sim 1$~GeV.

The function $H_K(Q^2)$ in Eq.(\ref{FiK}), like $H_{\Lambda}(Q^2)$ 
in Eq.(\ref{FiLam}), reflects the possible $Q^2$ dependence in the 
interaction vertex in Fig.1(b).
The simplest approximation is to assume point-like $\gamma^* K$ and 
$\gamma^* \Lambda$ couplings, $H_K(Q^2) = H_{\Lambda}(Q^2) = 1$\ 
\cite{MB}.
Of course the virtual $K$ and $\Lambda$ do have finite size, and their
coupling to $\gamma^*$ should in reality exhibit some $Q^2$ dependence.
The most natural way to model this $Q^2$ dependence is through the
vector meson dominance model, in which the virtual photon at low $Q^2$
couples to the kaon or $\Lambda$ through its fluctuations into correlated
$q\overline q$ pairs \cite{MD,JAF,HKW,HILMAR,HMD}.
Since the photon has $J^{PC} = 1^{- -}$, the states with the correct
quantum numbers to which the $\gamma^*$ can couple are the vector mesons,
$\rho^0$, $\omega$ and $\phi$.
Indeed, the $\omega$ and $\phi$ mesons can act as an extra source of
strangeness in the nucleon through this mechanism.
A detailed treatment of the vector meson dominance model in strange
form factors was presented in Refs.\cite{MD,JAF,HILMAR,HMD}, where 
it was found that $r_s^2$ increased by a factor $\sim 2$ with respect 
to the kaon loop only result ($\mu_s$ is of course not affected by any 
changes in the $Q^2$ dependence).
Other, quark-type models have also been used to estimate the intrinsic
form factors of the bare particles \cite{MB,ITO,CBM}. 
However, as outlined in Sec.I, our aim here is more to explore the 
relationship between strange quark distributions in coordinate and 
momentum space within a single specific model, rather than present
a comprehensive analysis of the form factors and structure functions
themselves.
Therefore in order to avoid diluting the result from the kaon cloud model
by introducing extra degrees of freedom into the calculation, for our
purposes we will begin by investigating the contributions to the $Q^2$ 
dependence of the strange form factors arising from kaon loops alone.

The $Q^2$ dependence of the strange Sachs electric and magnetic $G_E^s$ 
and $G_M^s$ form factors is shown in Figs.2 and 3 for cut-off masses 
$\Lambda_{K\Lambda} = 0.7, 1$ and $1.3$ GeV.
The values of the masses and couplings used in the numerical calculations 
are
$M = 939$ MeV,
$M_\Lambda = 1116$ MeV,
$M_\Sigma = 1190$ MeV,
$m_K = 494$ MeV and
$g_{KN\Lambda} = -13.98$,
$g_{KN\Sigma}  = 2.69$ \cite{HOLZ}.
The results in Figs.2 \& 3 contain contributions from both $K\Lambda$ 
and $K\Sigma$ components, with the latter contributing $\sim 4\%$ of the 
total, which reflects the ratio $g_{KN\Sigma}/g_{KN\Lambda} \sim -1/5$ 
expected from SU(3) symmetry.
Contributions from $KY^*$ loops to the strange form factors, where $Y^*$
represents the $J=3/2$ decouplet hyperons, are suppressed by more than 
two orders of magnitude relative to the $K\Lambda$ loops.
The sign of $G_E^s$ is positive, while that of $G_M^s$ negative, 
in agreement with the instant-form kaon cloud model results of 
Refs.\cite{MB,MD}.
The magnitude depends strongly on $\Lambda_{K\Lambda}$, although
even for the largest cut-off of $1.3$ GeV it is still somewhat 
smaller than in Refs.\cite{MB,MD}.
The strangeness (Sachs) radius is found to be small and negative, 
ranging from 
$r_s^2 \approx -0.004$ fm$^2$ for $\Lambda_{K\Lambda}=0.7$ GeV to 
$r_s^2 \approx -0.008$ fm$^2$ for $\Lambda_{K\Lambda}=1.3$ GeV.
This is slightly smaller than that obtained in previous kaon cloud
model calculations \cite{MB,MD,ITO}, although still within the same 
order of magnitude.
The strange Dirac radius is somewhat smaller still, 
$r_{s, Dirac}^s \approx -0.0007 \rightarrow -0.0012$ fm$^2$
for the two cut-off masses.
The strange magnetic moment is also found to be negative as in 
earlier studies \cite{MB,MD} although somewhat smaller, 
$\mu_s \approx -0.04 \rightarrow -0.1$.
To obtain the same numerical values for the form factors and radii as 
in Refs.\cite{MB,MD,ITO} would require values for the vertex function 
cut-off of $\Lambda_{K\Lambda} \sim 3$ GeV, as seen in Fig.4, where the 
$\Lambda_{K\Lambda}$ dependence of $r_s^2$, $r_{s,Dirac}^2$ and $\mu_s$ 
is plotted (note that $\mu_s$ is scaled down by a factor 10). 
For such hard vertex functions, however, one would find it difficult
to be consistent with the data on the $s$--$\overline s$ asymmetry,
as we shall discuss in the next Section.
Furthermore, the average number of kaons in the nucleon for such a 
cut-off mass would be $\langle n \rangle_{K\Lambda} \approx 0.3$, 
which is clearly beyond the limits of a perturbative treatment
of the meson cloud.

The matrix elements of the strange axial vector current 
$J_{5\mu}^s = \overline s \gamma_{\mu} \gamma_5 s$ are
parametrized in terms of the axial form factors $G_A$: 
and $G_P$:
\begin{eqnarray}
\langle N(P') | J_{5\mu}^s(0) | N(P) \rangle
&=& \overline u(P')
\left( 
\gamma_{\mu} \gamma_5 G_A^s(Q^2)
     + {q_{\mu} \over M} \gamma_5 G_P^s(Q^2)
\right) 
u(P).
\end{eqnarray}
The strange pseudoscalar form factor $G_P^s$ is not observable 
in semi-leptonic neutral current reactions \cite{MB}, and is 
therefore not considered here.
Because the kaon has spin 0, the strange axial vector form factor $G_A^s$ 
receives contributions only from the $\gamma^* \Lambda$ coupling in 
Fig.1(a).
In this case one has: 
\begin{eqnarray}
G_A^s(Q^2) 
&=& Q_{\Lambda} \int_0^1 dy\ \Delta f^{(\Lambda)}(y,Q^2)\ H_{\Lambda}(Q^2), 
\end{eqnarray}
where the light-cone axial distribution function $\Delta f^{(\Lambda)}$ 
is given by:
\begin{eqnarray}
\label{DfL}
\Delta f^{(\Lambda)}(y,Q^2)
&=& { g_{KN\Lambda}^2 \over 16 \pi^3 }
\int { d^2{\bf k}_\perp \over y^2 (1-y) }
{ {\cal F}({\cal M}^2_{\Lambda K,i})\
  {\cal F}({\cal M}^2_{\Lambda K,f})
  \over ({\cal M}^2_{\Lambda K,i} - M^2)
	({\cal M}^2_{\Lambda K,f} - M^2) }
\nonumber \\
& & \times
\left(
 - k_\perp^2 + (M_{\Lambda} - y M)^2
 + (1-y)^2 { q_\perp^2 \over 4 }
\right). 
\end{eqnarray}
The $Q^2$ dependence of the axial $G_A^s$ form factor is shown in Fig.5
for three values of the $KN\Lambda$ vertex function cut-off, 
$\Lambda_{K\Lambda} = 0.7, 1$ and $1.3$ GeV.
At zero transferred momentum the ratio
\begin{eqnarray}
\eta_s &=& {G_A^s(0) \over g_A},
\end{eqnarray}
where $g_A = 1.26$ from nucleon $\beta$-decays, measures the strange 
isovector axial charge of the proton.
For $KN\Lambda$ vertex function factor cut-offs between  
$\Lambda_{K\Lambda} = 0.7$ and $1.3$ GeV, $\eta_s$ is found to be 
very small, ranging between $\eta_s \approx 0.0017$ and 0.002.
Note that the sign of $\eta_s$ (and $G_A^s$) is {\em positive}, becoming 
negative only for $\Lambda_{K\Lambda} \agt 1.6$ GeV, as seen in Fig.5.
As we discuss in more detail in Sec.IV B, this fact illustrates that 
kaon loops alone cannot explain the apparent large strange component 
of the proton spin.

\section{Deep-Inelastic Scattering}

The meson cloud model of the nucleon has been successful in providing 
understanding of the origin of some of the symmetry breaking amongst 
the proton's sea quark distributions observed in recent experiments.
A pion cloud, for example, naturally allows one to account for the 
excess of $\overline d$ quarks over $\overline u$ in the proton 
\cite{NMC}.
In a similar vein, the kaon cloud of the nucleon gives rise to the 
observed SU(3) flavor symmetry breaking in the proton sea \cite{AWT}.
Furthermore, it also leads to different strange and antistrange quark
distributions in the nucleon, as first pointed out by Signal and Thomas
\cite{ST}.

This question has recently come to prominence again with the availability 
of new neutrino and antineutrino deep-inelastic scattering data, which 
were analyzed for a possible non-zero $s$--$\overline s$ difference 
\cite{CCFR}.
Such an asymmetry arises naturally in a kaon cloud picture of the nucleon, 
since the $s$ and $\overline s$ quarks have quite different origins in 
this model.
Indeed, it has been argued \cite{JI,ST} that because the $s$ quark comes
from the $\Lambda$, its distribution should be valence-like, while the 
$\overline s$, originating in the lighter kaon, should be much softer 
and resemble a typical sea distribution. 
On the other hand, the experimental $s/\overline s$ ratio was found 
to be consistent, within large errors, with unity, 
$s/\overline s \propto (1-x)^{-0.46 \pm 0.85 \pm 0.17}$, 
prompting suggestions \cite{JI} that the meson cloud model 
is ruled out by these data.

In this Section we discuss how the above argument is modified when one
takes into account the different $K$ and $\Lambda$ distribution functions
in the nucleon.
The difference $s$--$\overline s$ turns out to be very sensitive to the 
details of the hadronic vertex functions used in calculating these
distributions.
When calculated consistently in terms of the light-cone vertex functions
discussed in the previous Sections, the kaon cloud model predicts a very 
small excess of $\overline s$ over $s$ at large $x$, which is not in 
contradiction with the neutrino deep-inelastic data \cite{CCFR}.

\subsection{$s$--$\overline s$ Asymmetry}

Within the same impulse approximation in which the form factors in Sec.III
are calculated, the deep-inelastic quark distribution (at a scale $\mu^2$) 
in the meson cloud model can be written as a one-dimensional convolution 
of the meson or hyperon light-cone distribution function and the intrinsic
quark distribution in the meson or hyperon.
The $s$ and $\overline s$ quark distributions, generated from Figs.1(a) and 
(b), respectively, can be written \cite{ST,MTV}:
\begin{mathletters}
\label{conv}
\begin{eqnarray}
s(x,\mu^2)
&=& \int_x^1 { dy \over y } 
f_{\Lambda K}(y)\ s^{\Lambda}\left({x\over y},\mu^2\right),       \\ 
\overline s(x,\mu^2)
&=& \int_0^{1-x} { dy \over 1-y } 
f_{K\Lambda}(1-y)\ \overline s^{K^+}\left({x\over 1-y},\mu^2\right), 
\end{eqnarray}
\end{mathletters}%
where the $N \rightarrow K\Lambda$ splitting functions are related to 
the $\Lambda$ and $K$ distribution functions in Eqs.(\ref{f1L}) and
(\ref{f1K}) by:
\begin{mathletters}
\label{fDIS}
\begin{eqnarray}
f_{\Lambda K}(y)
&=& f_1^{(\Lambda)}(y,Q^2=0),	\\
f_{K\Lambda}(y)
&=& f_1^{(K)}(y,Q^2=0).
\end{eqnarray}
\end{mathletters}%
The net strangeness in the nucleon being zero implies that the 
distributions must be normalized such that: 
\begin{eqnarray}
\int_0^1 dx \left( s(x,\mu^2) - \overline s(x,\mu^2) \right)
&=& 0,
\end{eqnarray}
as can be explicitly verified from Eqs.(\ref{conv}).

As for the intrinsic form factors of the virtual $K$ and $\Lambda$, 
the intrinsic $K$ and $\Lambda$ structure functions are essentially 
unknown.
However, the advantage of the light-cone approach is that the intermediate
state particles are on-mass-shell, thus allowing the on-mass-shell structure
functions of the kaon and $\Lambda$ to be used \cite{MTV}. 
In a covariant instant-form formulation where the $K$ and $\Lambda$ are
off their mass shells, one needs to make additional assumptions about 
the extrapolation of the structure functions into the off-shell region,
and indeed about the definition of the structure function of an off-shell
$K$ or $\Lambda$ itself \cite{MST}.
According to the usual prescription adopted in the literature, in the 
instant-form calculations it is usually simply assumed that the 
off-mass-shell structure function is the same as that on-shell.
While possibly a reasonable approximation for heavy baryons, the
justification for such an ansatz is certainly not clear for the kaon, 
which is typically much further off-mass-shell.

The $\overline s$ distribution in kaons is obtained from measurements 
of final states in inclusive $K +$ target $\rightarrow V X$ reactions, 
where $V = \mu^+ \mu^-$ in Drell-Yan production \cite{K_DY}, or 
$V = \rho, \phi, \cdots$ in inclusive meson production \cite{K_V}. 
One finds that the ratio of the $K$ structure function to the much better
determined $\pi$ structure function \cite{PISF} is consistent with unity 
over most of the range of $x$, dropping slightly at large $x$, 
$q^K / q^{\pi} \sim (1-x)^{0.18 \pm 0.07}$ \cite{K_DY}.
%
%
For the $s$ quark distributions in $\Lambda$ and $\Sigma$ one can expect 
the SU(3) symmetric relations $s^{\Sigma^+} \sim d$ and 
$s^{\Sigma^0} \sim s^{\Lambda} \sim u/2$ to be reasonable approximations.

With these distribution functions, and the splitting functions in 
Eqs.(\ref{f12L}), (\ref{f12K}) and (\ref{fDIS}), the resulting 
$s$--$\overline s$ asymmetry is plotted in Fig.6 at $\mu^2 = 10$ GeV$^2$, 
for different values of the $KN\Lambda$ vertex function cut-off,
$\Lambda_{K\Lambda}=0.7, 1$ and $1.3$ GeV.
(Note that it would not be meaningful to compare the calculated $s$ and
$\overline s$ distributions in Eq.(\ref{conv}) separately with the 
structure function data, since the distributions calculated in the 
kaon cloud model do not contain any contributions from the perturbative 
process $g \rightarrow s\overline s$.)
The asymmetry in the kaon cloud model turns out to be very small, and 
for not too large values of $\Lambda_{K\Lambda}$, broadly consistent 
with the CCFR experiment within the given errors \cite{CCFR}.
To obtain the difference $s$--$\overline s$ we have used the absolute
values of $s + \overline s$ from the parametrizations of Refs.\cite{PARAM}.
We should point out, however, that there exists some controversy regarding
the overall normalization of the deep-inelastic neutrino data from which 
the strange quark distribution was extracted, resulting from an apparent 
inconsistency between the neutrino data and data on inclusive charm 
production \cite{GRV,BRMA}.
In addition, the CCFR data were collected with Fe nuclei targets, so that 
one needs to consider possible nuclear EMC corrections in the data analysis
\cite{BRMA} before making any definitive conclusions about the $s$ and
$\overline s$ distributions.
In view of these uncertainties in the data themselves, one can certainly
not conclude that the meson cloud model is inconsistent with the data 
from deep-inelastic scattering experiments.

The sign of the difference $s$--$\overline s$ is very sensitive to the 
vertex function used at the $KN\Lambda$ vertex.
With a vertex function that depends only on the virtuality of the 
off-mass-shell kaon, such as in Eq.(\ref{FFt}), the covariant perturbation
theory calculation \cite{JI,ST} leads to $s-\overline s > 0$ at large $x$.
The origin of this difference can be traced back to the fact that the 
light-cone $K$ distribution function peaks at larger values of $y$ compared 
with the instant-form distribution with the vertex function (\ref{FFt}), 
which is somewhat less symmetric about $y=1/2$, Fig.7.
Upon convoluting the more symmetric light-cone distribution with the
$s^{\Lambda}$ and $\overline s^K$ distributions, the $\sim (1-x)^3$
behavior of $s^{\Lambda}$ and the harder $\sim (1-x)$ behavior of
$\overline s^K$ are translated into a harder overall $\overline s$
distribution compared with the $s$.
On the other hand, since at small $y$ the instant-form distribution
$f^{(IF)}_{K\Lambda}(y) \gg f^{(LC)}_{K\Lambda}(y)$, the original $x$
dependence in $s^{\Lambda}$ and $\overline s^K$ is skewed to such an 
extent that the resulting $s$ distribution actually becomes dominant 
at large $x$.
As discussed in Sec.II, however, the instant-form function 
$f^{(IF)}_{K\Lambda}(y)$, evaluated with the $t$-dependent 
vertex function (\ref{FFt}), does not satisfy the probability 
conservation relation in Eq.(\ref{sym}).
Hence we believe that the results for the $s$--$\overline s$ 
difference in Fig.6 are more realistic for the light-cone 
distribution functions.

\subsection{Polarized Strangeness}

Given the simplicity with which one can describe symmetry breaking in the 
nucleon's sea quark distributions with the meson cloud model, it is tempting 
to attribute the large violation of the Ellis-Jaffe sum rule \cite{SPIN} 
to a kaon cloud of the nucleon.
By carrying away some orbital angular momentum of the nucleon, a kaon cloud
could naively be expected to reduce the intrinsic spin carried by quarks, 
and give rise to a negatively polarized strange quark distribution, 
$\Delta s(x,\mu^2)$.
The effect of the meson cloud on the polarized nucleon quark distributions
has been addressed by several authors \cite{ZOL,JUL,SHT}.
Although the kaon cloud picture can provide a simple framework within 
which the non-perturbative sea could be understood qualitatively, it 
is clearly important to determined whether such a picture can provide,
or even be consistent with, a more quantitative description.
In view of the sensitivity of the sign and magnitude of the strange form
factors and the unpolarized $s$--$\overline s$ difference to the hadronic 
vertex functions, we shall use the light-cone framework to re-examine the
question of what role kaons play in the proton spin problem.

The contribution to the polarized strange quark distribution in the nucleon
from kaon loops can be written: 
\begin{mathletters}
\label{Dconv}
\begin{eqnarray}
\Delta s(x,\mu^2) 
&=& \int_x^1 {dy \over y} 
\Delta f_{\Lambda K}(y)
\Delta s^{\Lambda}\left({x\over y},\mu^2\right),  \\
\Delta \overline s(x,\mu^2)
&=& 0, 
\end{eqnarray}
\end{mathletters}%
where $\Delta \overline s$ is zero for the same reason that the $G_A^s$ 
form factor receives contributions only from the $\gamma^* \Lambda$
interaction diagram in Fig.1(a).
The helicity-dependent $N \rightarrow K \Lambda$ splitting function
is given by:
\begin{eqnarray}
\label{Delf}
\Delta f_{\Lambda K}(y)
&=& \Delta f^{(\Lambda)}(y,Q^2=0),
\end{eqnarray}
with $\Delta f^{(\Lambda)}(y,Q^2)$ defined in Eq.(\ref{DfL}).

If one had available information about the spin-dependent $x$-distribution 
of the strange quark in the bare $\Lambda$, one could predict the 
resulting $x$-dependence of $\Delta s(x,\mu^2)$ arising from kaon loops.
Unfortunately, there is no information about polarized quark 
distributions in any hadron other than the nucleon.
Nevertheless, one can still get an estimate of the size of the kaon loop
contribution by considering the first moment of $\Delta s(x,\mu^2)$.
In the SU(6) symmetric model the spin of the $\Lambda$ is carried
entirely by the $s$ quark.
It may seem reasonable therefore to expect that 
$\Delta s^{\Lambda}(\mu^2)
\equiv \int_0^1 dx\ \Delta s^{\Lambda}(x,\mu^2) 
\sim {\cal O}(1)$ at some low scale $\mu^2$, even taking into account 
SU(6) symmetry-breaking effects, or some of the spin residing on gluons 
or in the form of angular momentum.
Evolution to larger values of $\mu^2$ would imply that 
$\Delta s^{\Lambda}(\mu^2) \alt {\cal O}(1)$ \cite{EVOL}.
In this case the total strange quark contribution to the proton spin 
would be:
\begin{eqnarray}
\Delta s(\mu^2) 
&\equiv& \int_0^1 dx\ \Delta s(x,\mu^2)\ 
\alt\ \int_0^1 dy\ \Delta f_{\Lambda K}(y)\ 
=\ G_A^s(0). 
\end{eqnarray}
For cut-offs $\Lambda_{K\Lambda} = 0.7$--1.3 GeV this would give 
$\Delta s \alt 0.002-0.003$.
This is to be compared with values $\Delta s \sim -0.1$ quoted in 
connection with determinations of $\Delta s$ from the spin-dependent 
structure functions in Ref.\cite{SPIN}.
These results therefore suggest that a kaon cloud alone cannot be 
expected to reproduce the observed deviation from the Ellis-Jaffe
sum rule.
Other mechanisms for generating a non-zero $\Delta s$, such as ones
based on the gluon axial U(1) anomaly, need to be invoked to account 
for the data.

\section{Conclusion}

We have presented a framework for studying strange form factors and 
quark distributions of the nucleon consistently in terms of the meson 
cloud model.
Working on the light-cone, one avoids many of the problems and ambiguities
associated with the choice of momentum dependent hadronic vertex functions 
encountered in instant-form calculations.

Within the uncertainty range of the input parameters, the strangeness 
(Sachs) radius is found to be very small and negative, in the vicinity 
	$r_s^2 \approx -0.004 \rightarrow -0.008$ fm$^2$ 
for $KN\Lambda$ vertex function cut-offs of 
$\Lambda_{K\Lambda}=0.7$--$1.3$ GeV.
The strange Dirac radius is somewhat smaller, 
	$r_{s,Dirac}^s \approx -0.0007 \rightarrow -0.0012$ fm$^2$.
The strange magnetic moment is found to be 
	$\mu_s \approx -0.04 \rightarrow -0.1$, 
which is two to three times smaller than in previous estimates.
The strange axial vector charge is also small, 
	$\eta_s \approx 0.0017 \rightarrow 0.002$, 
but in addition comes with the opposite sign compared with instant-form 
calculations with the meson cloud model.
Combined, these results suggest that the strangeness content of the 
nucleon at low $Q^2$ is indeed very small, and may represent a challenge
to experimentalists seeking to verify its presence.

Using the same light-cone framework, we have estimated the asymmetry
between the $s$ and $\overline s$ quark distributions in the nucleon,
which has been the subject of recent experimental investigation 
\cite{CCFR,GRV}.
We find the magnitude of the $s$--$\overline s$ difference to be very 
small, with the $\overline s$ distribution slightly harder than the $s$,
but somewhat sensitive to the shape of the hadronic $KN\Lambda$ vertex 
function.
Within the current experimental errors this is consistent with the recent
experimental determination of the asymmetry from neutrino deep-inelastic 
scattering, if moderately soft $KN\Lambda$ vertex functions are used.
For very hard vertex functions, the predicted asymmetry seems to be 
somewhat large in comparison with the data of Ref.\cite{CCFR}.
However, to be more definitive, more statistics on the charm production
data are needed, and the apparent discrepancy between the inclusive 
deep-inelastic muon and neutrino data and those on $c\bar c$ production
must be resolved \cite{CCFR,GRV,BRMA}.
Finally, the contribution of the kaon cloud to the polarized $\Delta s$ 
quark distribution is very small and {\em positive}, 
	$\Delta s \alt 0.002 \rightarrow 0.003$,
and therefore is unlikely to feature prominently in any final explanation
of the proton spin puzzle.

Aside from the experimental uncertainties, the small values for both 
the slope of the strangeness form factor and the $s$--$\overline s$ 
difference within the simple model considered here gives some support
to the suggestion of Ref.\cite{JI} that the coordinate and momentum 
space strange quark distributions are correlated.
This relationship will be clarified when new, more precise data from
deep-inelastic neutrino scattering become available.
Measurement of the $Q^2$ dependence of strange matrix elements at low 
and intermediate values of $Q^2$ at MIT-Bates, Jefferson Lab and 
elsewhere will also be very valuable \cite{BATES,JEFF}.
The dynamical source of intrinsic strangeness can also be explored 
in semi-inclusive experiments involving coincidence measurement of 
scattered leptons and specific hadronic final states \cite{ZPA}.
Measurement of the spin transfer from target protons to recoiling $\Lambda$ 
hyperons in the target fragmentation region could discriminate between kaon
cloud models, in which the spins are highly correlated, and parton 
fragmentation models, for which the correlations are very weak \cite{ZPA}.
Similar effects have also been discussed in hadronic reactions, 
$\overline p p \rightarrow \overline \Lambda \Lambda$ \cite{AEK}.
While the upcoming experiments will be difficult, if successful they 
should provide quite important information on the role of strangeness 
in the nucleon, as well as its dynamical origin.

\acknowledgements

We would like to thank M.K. Banerjee for useful remarks and suggestions. 
Helpful discussions with M. Burkardt, T.D. Cohen, H. Ito, X. Ji, B.Q. Ma,
M. Nielsen, J. Speth, F.M. Steffens, A.W. Thomas and S. Wang
are gratefully acknowledged.
MM would like to thank the TQHN group at the University of Maryland for 
their hospitality during his extended visit, and the Brazilian agency
CAPES for the financial support which made this visit possible.
This work was supported by the Department of Energy grant DE-FG02-93ER-40762.


\begin{figure}
\epsfxsize=15cm 
\epsffile{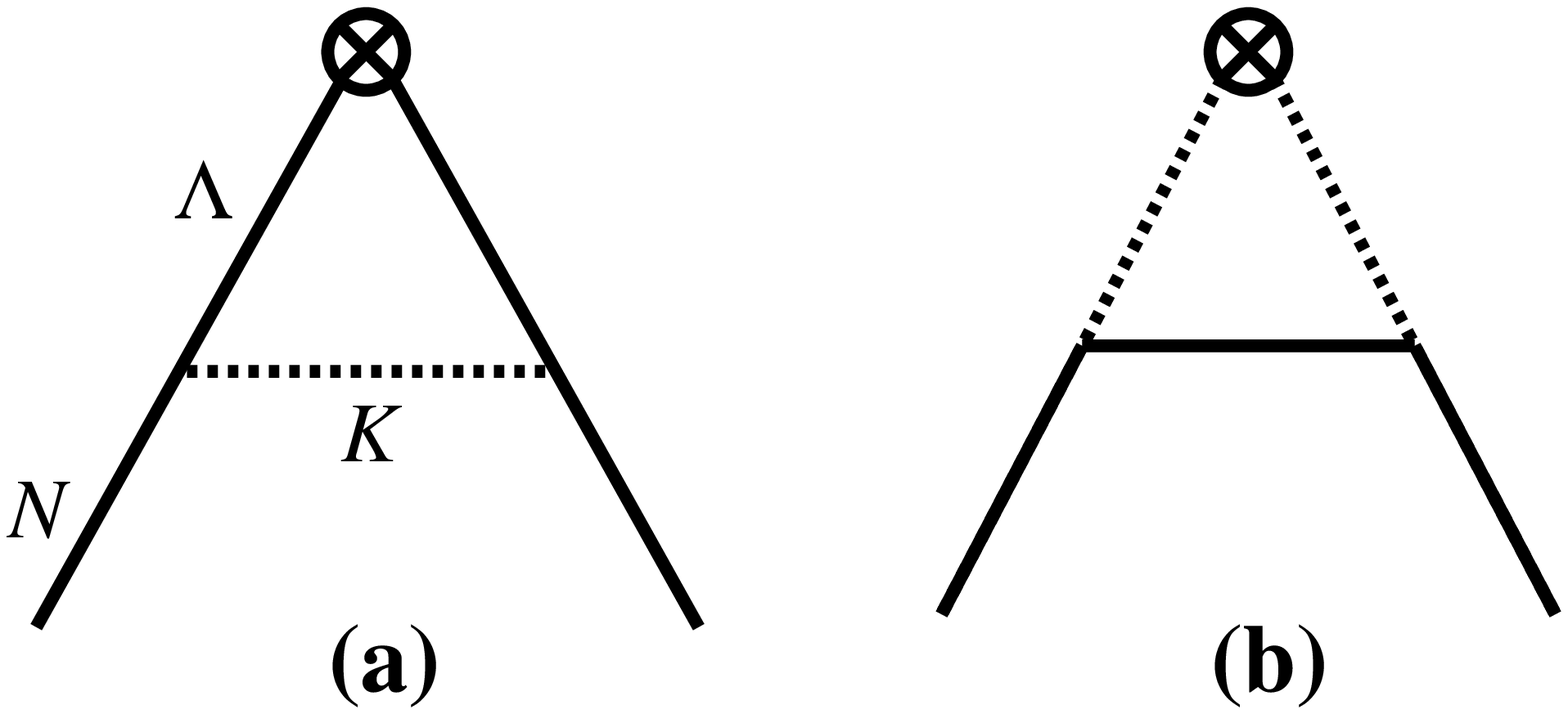}
\caption{Kaon cloud model of the nucleon: interaction of the probe with 
	the (a) hyperon ($``\Lambda''$), (b) kaon.}
\end{figure}

\begin{figure}
\epsffile{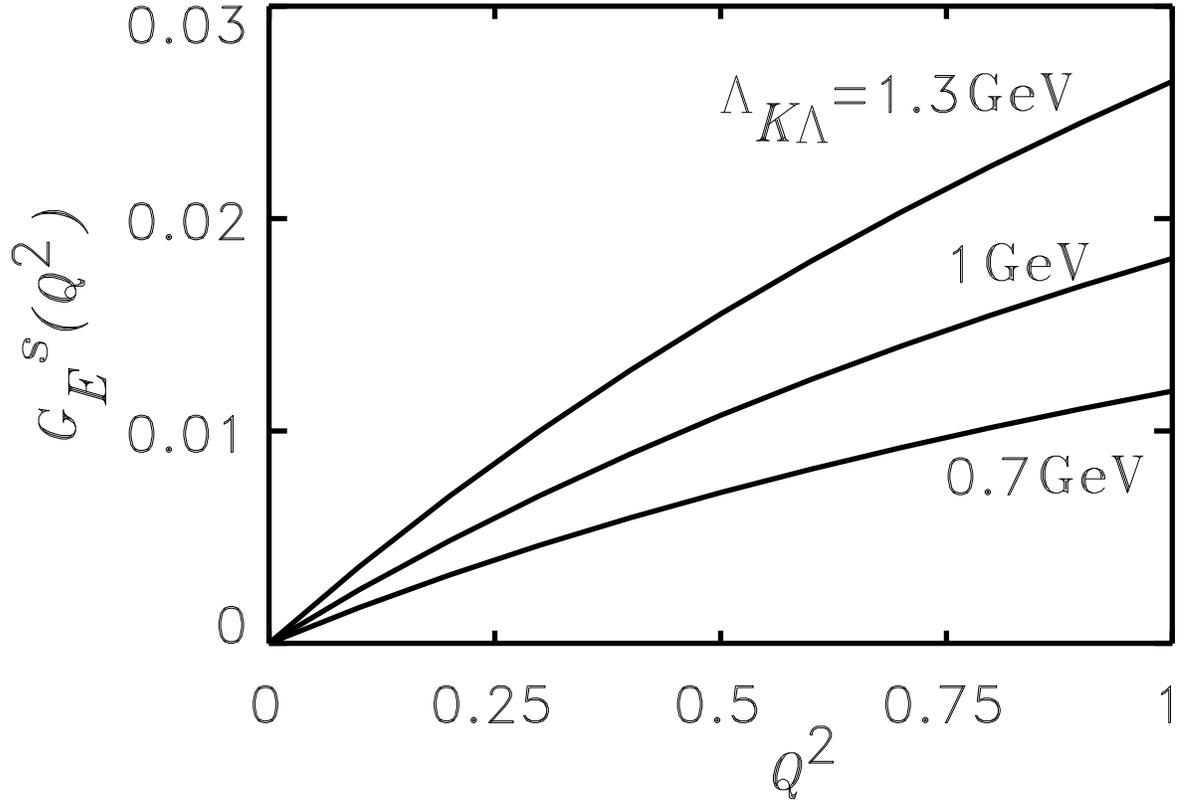}
\caption{$Q^2$ dependence of the Sachs electric strange form factor of 
	the nucleon, $G_E^s$, for various $KN\Lambda$ vertex function 
	momentum cut-offs.  Contributions from $K\Sigma$ components are
	also included.}
\end{figure}

\begin{figure}
\epsffile{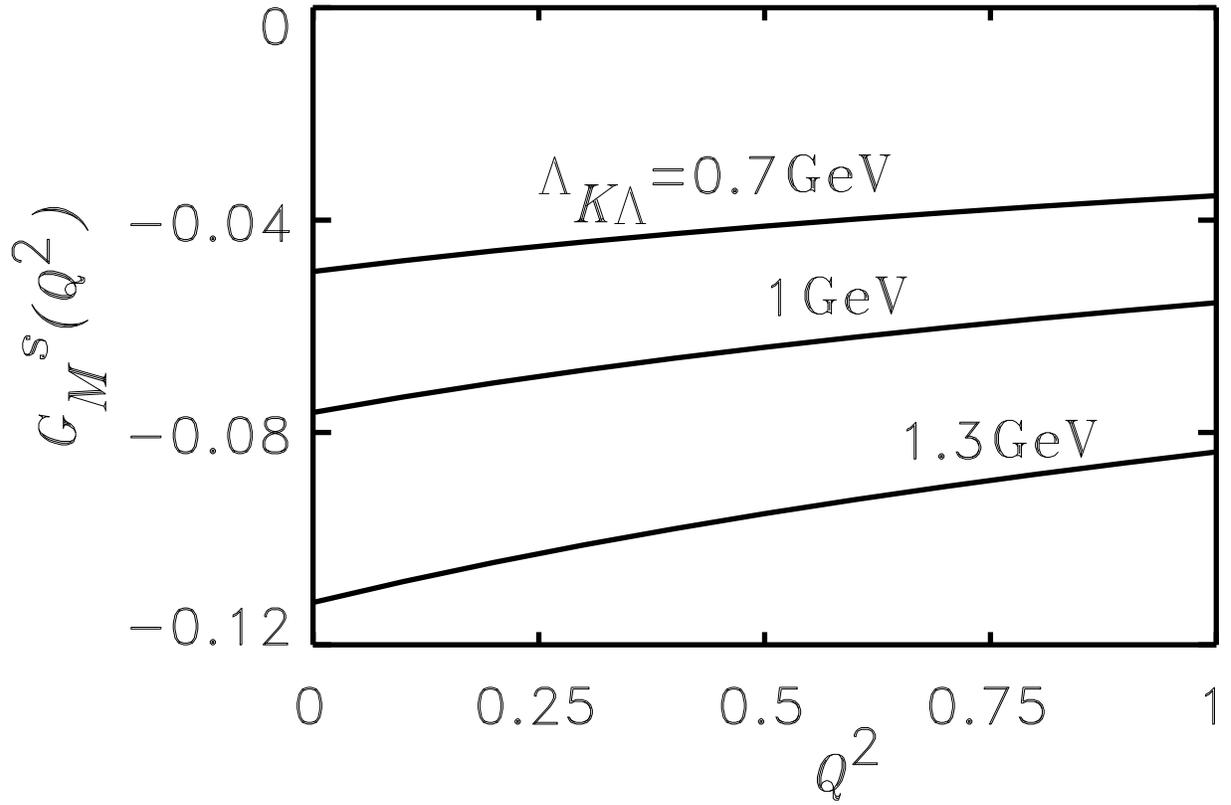}
\caption{$Q^2$ dependence of the Sachs magnetic strange form factor of 
	the nucleon, $G_M^s$.}
\end{figure}

\begin{figure}
\epsffile{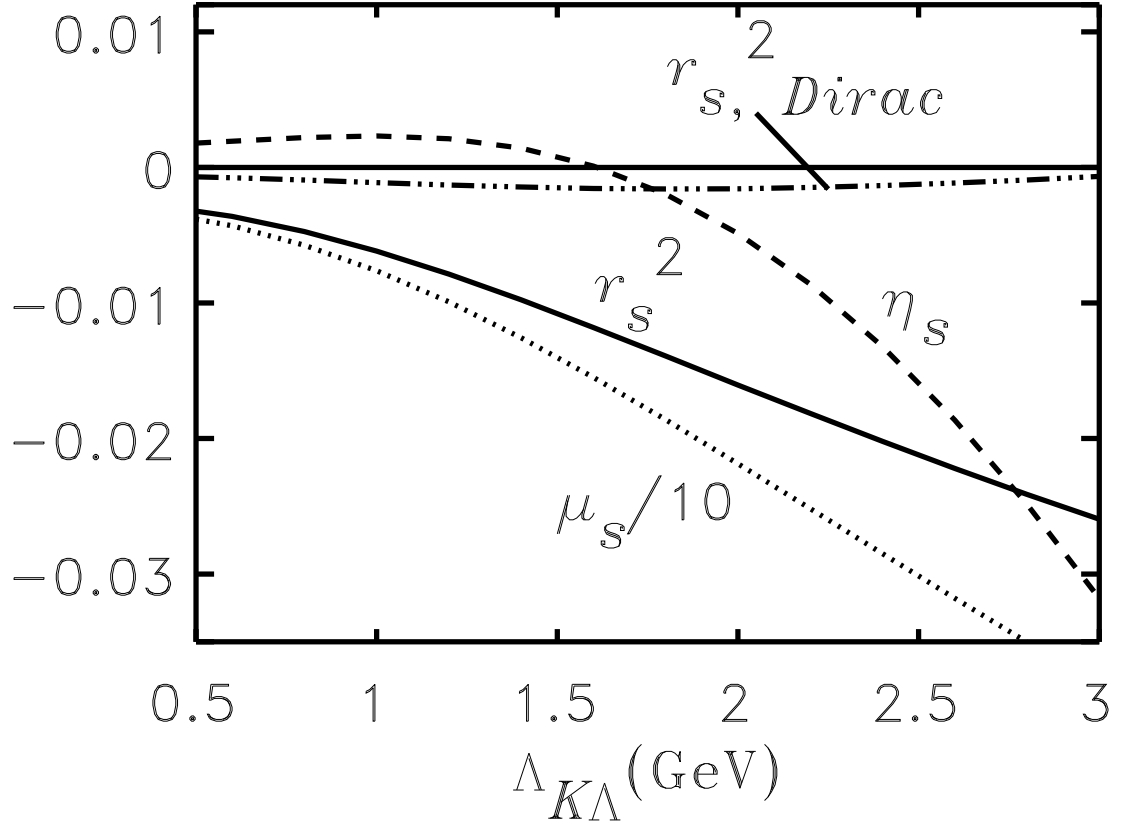}
\caption{Cut-off dependence of the strangeness radii, $r_s^2$ and
	$r_{s, Dirac}^2$ (in fm$^2$), strange magnetic moment,
	$\mu_s$ (scaled by 1/10), and strange axial charge, $\eta_s$, 
	of the nucleon.}
\end{figure}

\begin{figure}
\epsffile{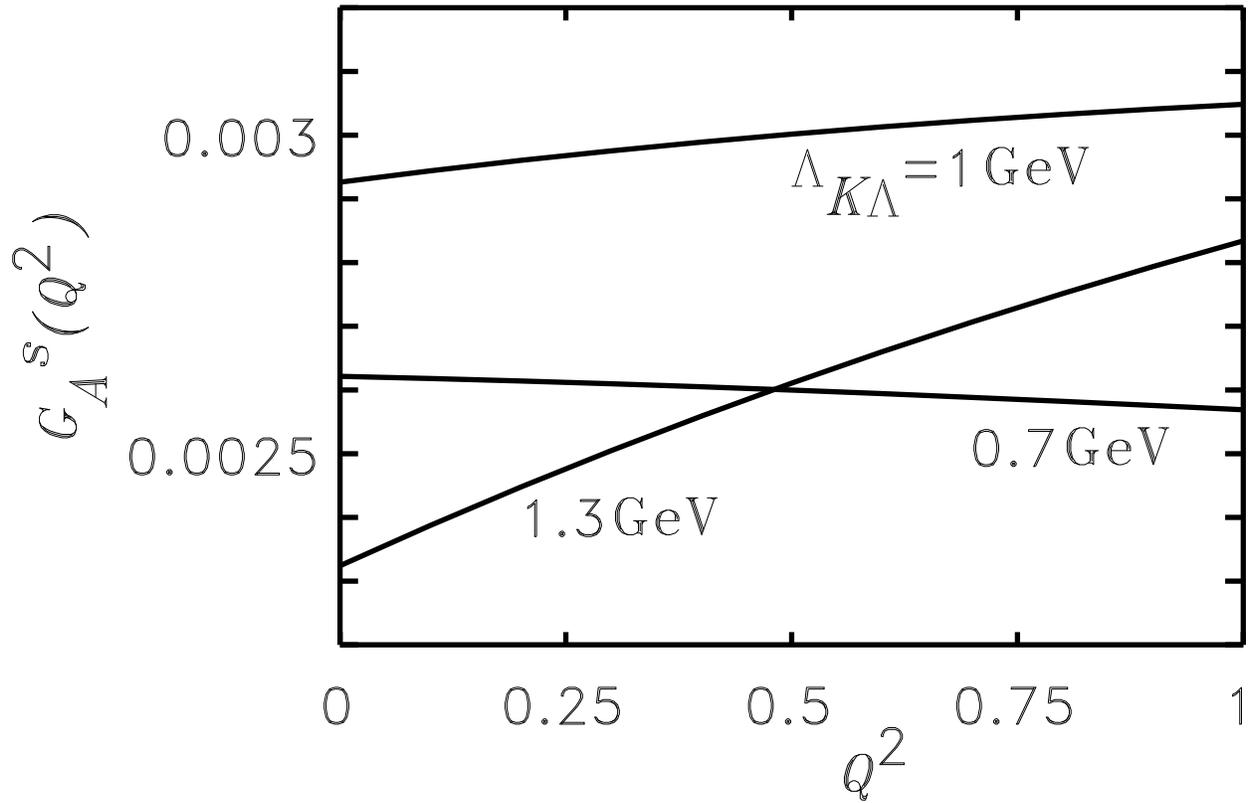}
\caption{$Q^2$ dependence of the strange axial vector form factor of
        the nucleon, $G_A^s$.}
\end{figure}

\begin{figure} \epsffile{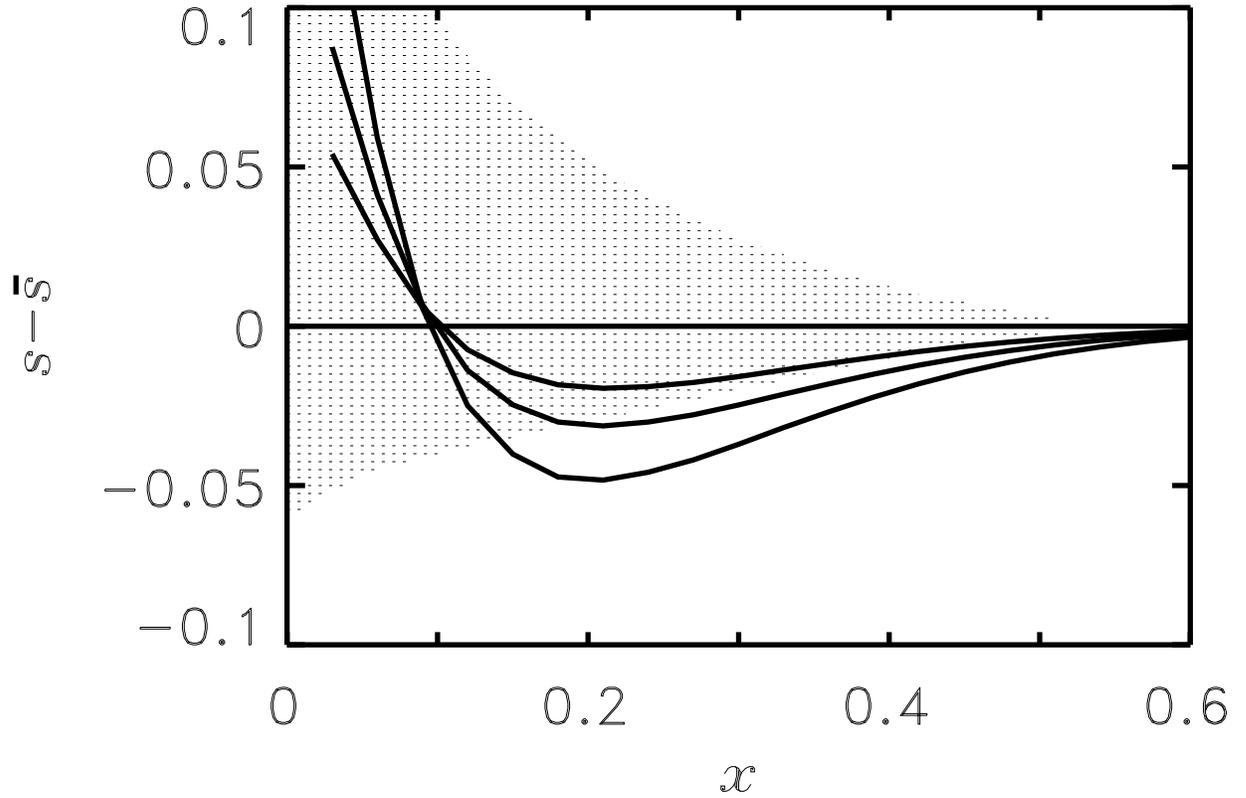}
\caption{Strange--antistrange quark distribution asymmetry in the nucleon.
	The solid lines correspond to the asymmetry calculated for 
	$\Lambda_{K\Lambda} = 0.7$ GeV (smallest asymmetry), 1 GeV and
	1.3 GeV (largest asymmetry), while the shaded region represents
	the uncertainty range of the data \protect\cite{CCFR}.} 
\end{figure}

\begin{figure}
\epsffile{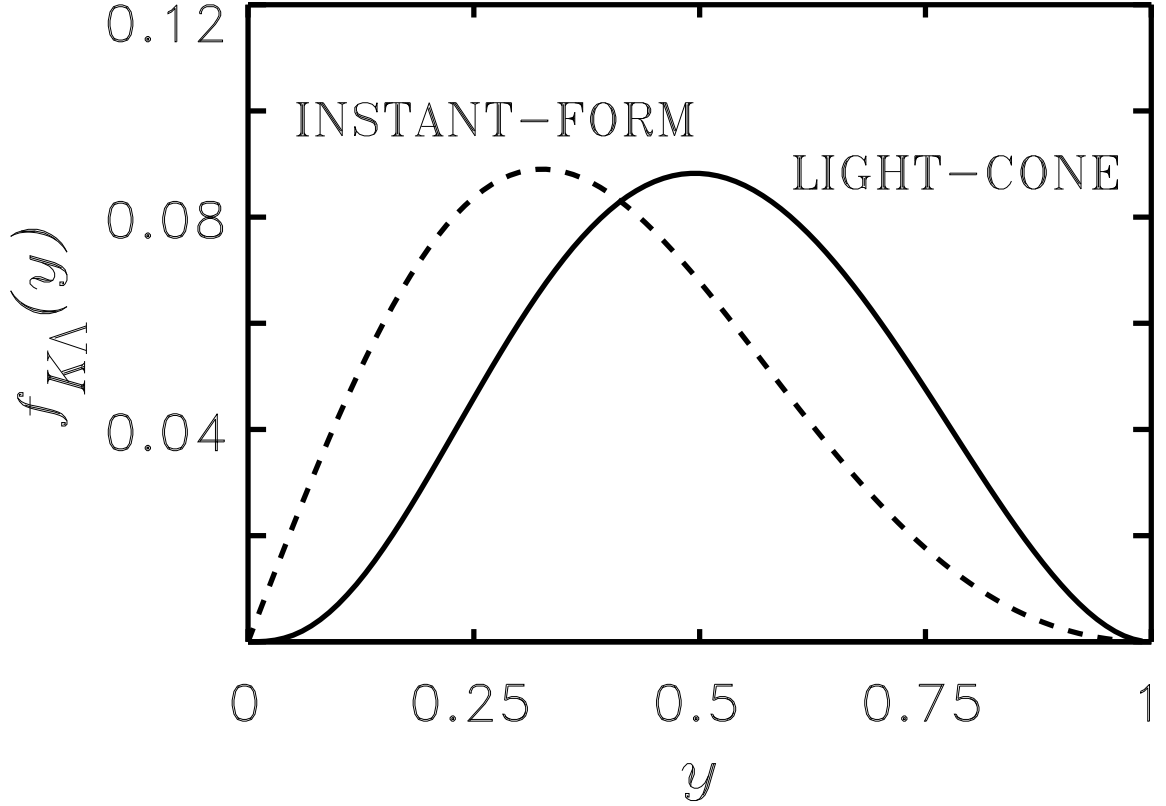}
\caption{Splitting function for $N \rightarrow K\Lambda$, $f_{K\Lambda}(y)$.
	The solid curve is the light-cone distribution function,
	for a vertex function cut-off mass $\Lambda_{K\Lambda}=1$ GeV;
	dashed is the instant-form result with the $t$-dependent 
	form factor in Eq.(\protect\ref{FFt}), which does not satisfy
	Eq.(\protect\ref{sym}), normalized to give the same value for
	$\langle n \rangle_{K\Lambda}.$}
\end{figure}

\end{document}